%% file: pwv_paper.tex
\documentclass[twocolumn]{aastex61}
\usepackage{color}
\usepackage{listings}
\usepackage{url}
\usepackage{verbatim}
\usepackage{subfigure}
\newcommand{\code}[1]{\texttt{#1}}
\newcommand{\pwvkpno}{\texttt{pwv\_kpno}}

\definecolor{dkgreen}{rgb}{0,0.6,0}
\definecolor{gray}{rgb}{0.5,0.5,0.5}
\definecolor{mauve}{rgb}{0.58,0,0.82}
\submitjournal{PASP}
\shorttitle{PWV Atmospheric Transmission Models in Python}
\shortauthors{Perrefort, Wood-Vasey et al.}

\begin{document}

\title{pwv\_kpno: A Python Package for Modeling the Atmospheric Transmission Function due to Precipitable Water Vapor}

\correspondingauthor{Daniel Perrefort}
\email{djperrefort@pitt.edu}
\keywords{atmospheric effects; methods: observational; techniques: image processing}

\author{Daniel Perrefort}
\affiliation{
    Pittsburgh Particle Physics, Astrophysics, and Cosmology Center (PITT PACC).
    Department of Physics and Astronomy,
    University of Pittsburgh,
    Pittsburgh, PA 15260, USA
    }
\affiliation{   
    Visiting astronomer, Kitt Peak National Observatory, National Optical Astronomy Observatory,
    which is operated by the Association of Universities for Research in Astronomy (AURA) under 
    a cooperative agreement with the National Science Foundation.
    }

\author{W.~M. Wood-Vasey}
\affiliation{
    Pittsburgh Particle Physics, Astrophysics, and Cosmology Center (PITT PACC).
    Department of Physics and Astronomy,
    University of Pittsburgh,
    Pittsburgh, PA 15260, USA
    }
\affiliation{   
    Visiting astronomer, Kitt Peak National Observatory, National Optical Astronomy Observatory,
    which is operated by the Association of Universities for Research in Astronomy (AURA) under 
    a cooperative agreement with the National Science Foundation.
    }

\author{K. Azalee Bostroem}
\affiliation{
    Department of Physics,
    University of California,
    Davis, CA 95616, USA
    }

\author{Kirk Gilmore}
\affiliation{
    SLAC National Accelerator Laboratory,
    Menlo Park, CA 94025, USA
    }

\author{Richard Joyce}
\affiliation{
    National Optical Astronomy Observatory,
    Tucson, AZ 85719, USA
    }
    
\author{Tom Matheson}
\affiliation{
    National Optical Astronomy Observatory,
    Tucson, AZ 85719, USA
    }

\author{Charles Corson}
\affiliation{
    National Optical Astronomy Observatory,
    Tucson, AZ 85719, USA
    }

\input{abstract}
\input{introduction}
\input{background}
\input{features}
\input{other_locations}

\input{validation}

\input{demonstration}
\input{conclusion}
\input{acknowledgments}

\bibliographystyle{aasjournal}
\bibliography{pwv_paper}

\end{document}

%% file: abstract.tex
\begin{abstract}
    We present a Python package, \pwvkpno, that provides models for the atmospheric transmission due to precipitable
    water vapor (PWV) at user specified sites. Using the package, ground-based photometric observations taken
    between $3,000$ and $12,000$ \r A can be corrected for atmospheric effects due to PWV. Atmospheric 
    transmission in the optical and near-infrared is highly dependent on the PWV column density along the line of sight.
    By measuring the delay of dual-band GPS signals through the atmosphere, the SuomiNet project provides accurate PWV 
    measurements for hundreds of locations around the world. The \pwvkpno\ package uses published SuomiNet data in 
    conjunction with MODTRAN models to determine the modeled, time-dependent atmospheric transmission. 
    A dual-band GPS system was installed at Kitt Peak National Observatory (KPNO) in the spring of 2015. Using 
    measurements from this receiver we demonstrate that we can successfully predict the PWV at KPNO from nearby 
    dual-band GPS stations on the surrounding desert floor. The \pwvkpno\ package can thus provide atmospheric
    transmission functions for observations taken before the KPNO receiver was installed. Using PWV measurements
    from the desert floor, we correctly model PWV absorption features present in spectra taken at KPNO. 
    We also demonstrate how to configure the package for use at other observatories.
\end{abstract}

%% file: introduction.tex
\section{Introduction} \label{sec:introduction}
    Upcoming
    ground-based surveys, such as  the Large Synoptic Survey Telescope,
    will require a photometric precision of one percent or better. Understanding
    and calibrating for the effects of atmospheric absorption is an important part
    of achieving this precision level (see \cite{li16}, \cite{burke14}, and
    \cite{burke10}). Ground-based photometry redward of 5,500~\r A suffers from
    significant and variable opacity due to water vapor in the atmosphere. 
    While ozone and aerosol scattering also play significant roles, their
    opacity is relatively smooth with wavelength.  In contrast, the absorption
    due to precipitable water vapor (PWV) has a distinct and complex spectrum. 
        
    Astronomers traditionally calibrate broad-band imaging by using a reference 
    catalog to compute correction terms for color, airmass, and perhaps a 
    higher-order color-airmass term. This approach implicitly accounts for the 
    effects of atmospheric opacity on observed images. In general, the color 
    term accounts for the difference in filter and detector sensitivity with 
    wavelength, but also includes some average contribution of the atmosphere 
    above the telescope being used.

    More detailed information can be obtained by observing a telluric standard 
    star.  These bright stars of known spectral energy distribution are well 
    suited for determining the absorption and scattering of the atmosphere.  In 
    order to describe atmospheric effects, spectroscopy should be performed on 
    a telluric standard at the same airmass as a desired target. This is 
    ideally performed at the same position and time as the photometric 
    observations. The total atmospheric absorption per wavelength can then be 
    found by dividing the observed spectrum by tabulated results already 
    corrected for absorption.

    While this method is effective, the majority of telescopes are not 
    configured to have an auxiliary spectrograph for observing telluric stars.  
    Because atmospheric absorption is variable over time, observations of a 
    standard star must be performed repeatedly and within a short time interval 
    of other targets. Even in setups with the capability to easily switch back 
    and forth between mosaic imaging and single-object spectroscopy, such 
    observations require diverting valuable observation time away from other 
    targets.

    As an alternative, astronomers commonly express the atmospheric absorption 
    as a linear function of airmass. Using photometric observations taken over 
    a range of airmass values, corrections are performed by fitting the linear 
    function in each band. This approach assumes that the absorption scales 
    linearly with airmass.  However, the absorption spectrum of water is a 
    complex series of very narrow absorption lines.  These individual lines can 
    saturate, and thus the absorption does not scale linearly with airmass.  
    This non-linearity introduces errors due to higher order effects when 
    calibrating photometric images \citep{blake11}.

    In the redder range of CCD sensitivity ($5,500 < \lambda < 12,000$ \r A), 
    the atmospheric transmission function is dominated by absorption due to 
    precipitable water vapor (PWV). The use of GPS to measure the localized, 
    PWV column density is a recently emerged technology in astronomy and an 
    accurate alternative to traditional methods \citep{dumont01}.  Through the 
    use of atmospheric modeling, these PWV measurements can be used to simulate 
    the atmospheric transmission. The resulting transmission function can then 
    be used to correct photometric observations of sources with known spectral 
    energy distributions for atmospheric absorption.

    We here introduce \pwvkpno\footnote{\pwvkpno\ can be downloaded using the 
    pip package manager or at \url{https://mwvgroup.github.io/pwv_kpno/}}: a 
    Python package that provides models for the atmospheric transmission due to 
    H$_2$O at user-specified sites. By using MODTRAN models 
    \citep{modtran} in conjunction with publicly available PWV measurements, 
    \pwvkpno\ is able to return models for the atmospheric transmission between 
    $3,000$ and $12,000$ \r A. 
    The package was beta tested at Kitt Peak National Observatory
    (KPNO) using a dual-band GPS system that was installed at the 
    WIYN 3.5-m telescope in 2015. Thus, direct PWV measurements at Kitt Peak are 
    available starting in early 2015, but by using measurements from stations 
    on the surrounding desert floor, the package is capable of modeling the 
    atmospheric transmission for years 2010 onward. The package also provides 
    access to tabulated PWV measurements, along with easy to use utility 
    functions for retrieving and processing newly published PWV data.

    In Section~\ref{sec:background} we discuss the use of PWV measurements as a 
    tool for correcting photometric observations. In 
    Section~\ref{sec:features_and_use} we present the features and 
    functionality of \pwvkpno, including how to access tabulated PWV 
    measurements and the package's modeling capabilities. Section 
    \ref{sec:other_locations} demonstrates how to model the atmosphere for a
    user-specified site other than Kitt Peak. In Section \ref{sec:validation} we
    present a validation of the package as a tool for correcting 
    ground-based observations. A demonstration 
    of how to use the package to correct for atmospheric effects is presented 
    in Section~\ref{sec:package_demo}. Finally, we present our conclusions in 
    Section~\ref{sec:conclusion}.
    \newline
    \newline

    \begin{figure*}
        \includegraphics[width=\textwidth]{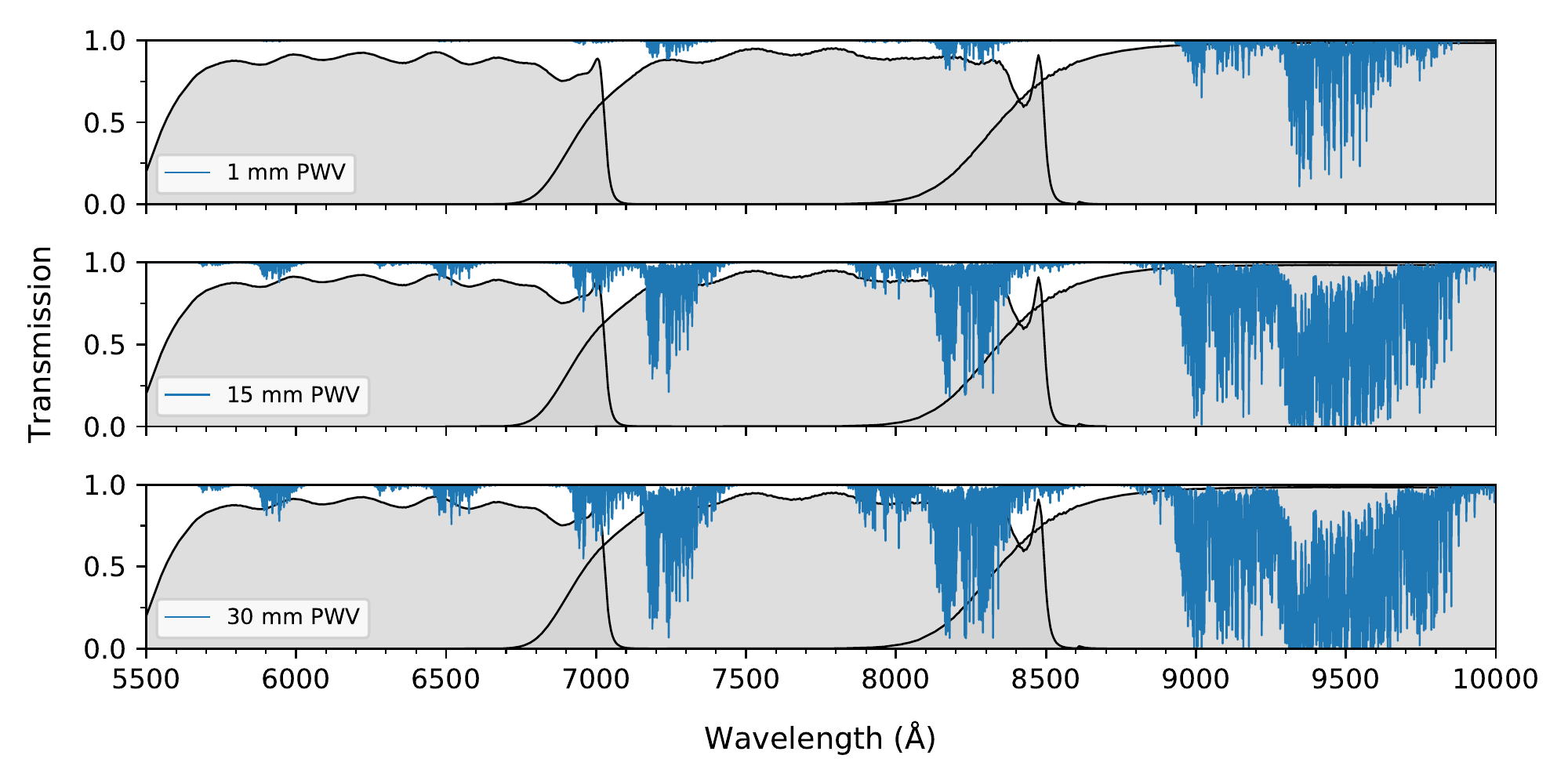}
        \caption{The $r$, $i$ and $z$ band mosaic filters of Kitt Peak
                 National Observatory (grey) compared against the MODTRAN 
                 modeled atmospheric transmission function due to precipitable 
                 water vapor (blue). Atmospheric transmission functions are 
                 shown for an airmass of one and a precipitable water vapor 
                 (PWV) column density of $1$ mm (top), $15$ mm (middle), and 
                 $30$ mm (bottom). Note that absorption features do not scale 
                 linearly with PWV, and some saturate at relatively low column 
                 densities.}
        \label{fig:pwv_comparison}
    \end{figure*}

    \begin{figure*}
        \includegraphics[width=\textwidth]{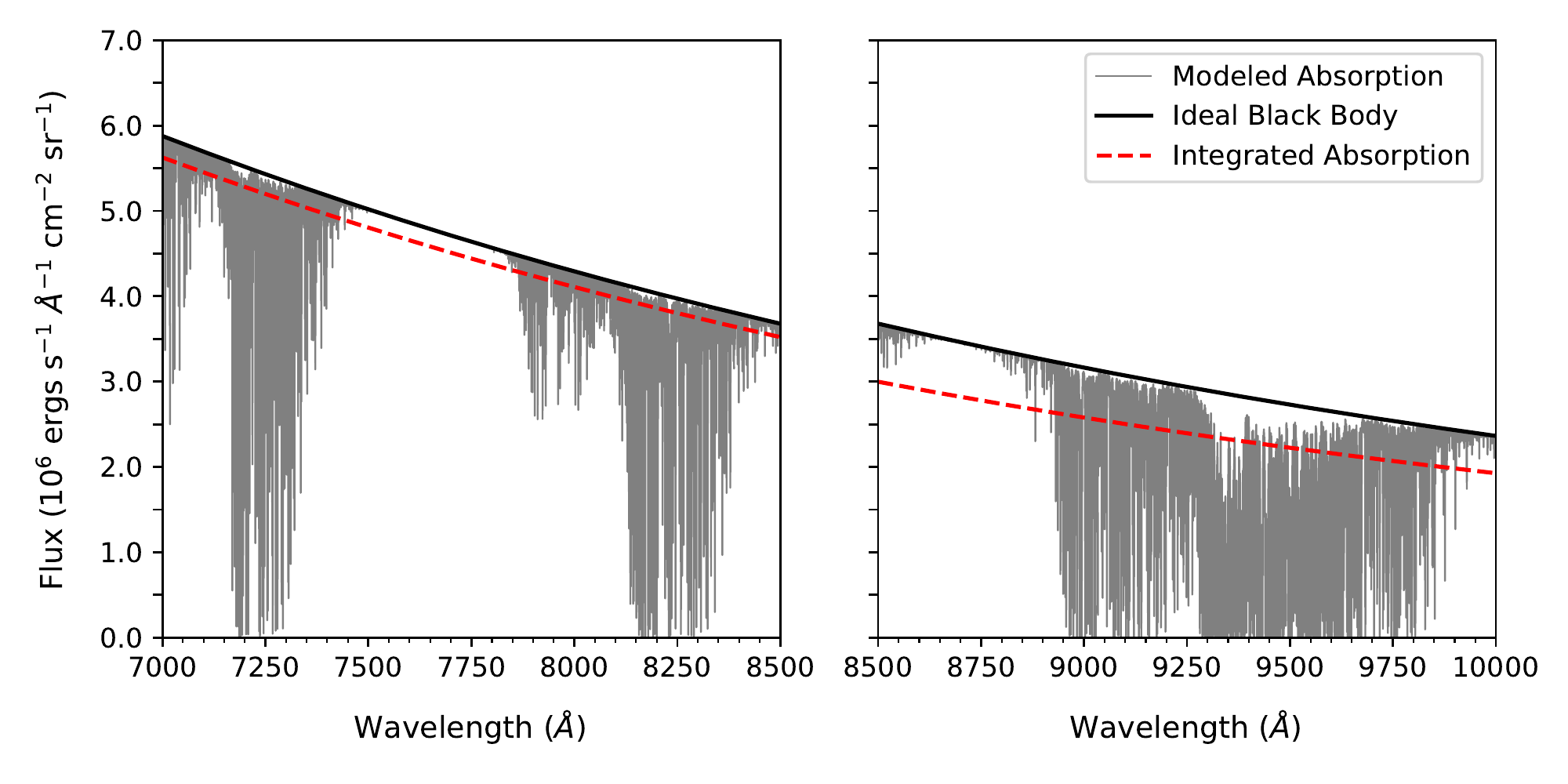}
        \caption{The SED of a blackbody at 8,000 K (black) across the $i$-band 
                 (left) and $z$-band (right) ranges. Shown in grey, the modeled 
                 atmospheric absorption for a PWV column density of 15 mm is 
                 applied to the SED. This is compared to the black body SED 
                 scaled using the integrated absorption in each band in red.}
        \label{fig:black_body}
    \end{figure*}

%% file: background.tex
\section{Background} \label{sec:background}
    The atmospheric transmission between $5,500$ and $12,000$ \r A is
    dominated by absorption due to PWV (Figure~\ref{fig:pwv_comparison}). The
    strength of PWV absorption lines in observed spectra correlate strongly
    with measurements of localized PWV column density \citep{blake11}. This 
    indicates that PWV measurements can be combined with atmospheric models to
    provide estimates of the atmospheric transmission at a given date and time.
    However, accomplishing this requires a source of accurate and readily
    accessible PWV measurements. Furthermore, since PWV levels can change by
    over 10\% per hour, measurements must be available in close to real time.

    By measuring the delay of dual-band GPS signals traveling through the
    atmosphere, it is possible to determine the PWV column density along the line
    of sight (see \cite{braun01}, \cite{dumont01}, and \cite{nahmias04}). This
    approach is made even more appealing by the existence of several established
    GPS networks dedicated to the measurement of geological and meteorological
    data on the international scale. The SuomiNet project \footnote{For more
    information see \url{https://www.suominet.ucar.edu}} \citep{ware00} is a 
    meteorological initiative that uses data from multiple GPS networks to
    provide semi-hourly PWV measurements. It currently publishes meteorological
    data from hundreds of receivers throughout the United States and Central 
    America.

    \subsection{Effects of PWV on Photometric Calibration} \label{ssec:calibration_effects}
    When correcting photometric observations for atmospheric effects, 
    astronomers commonly express atmospheric absorption as a linear 
    function of airmass. In this approach photometric observations are corrected 
    by fitting for a set of extinction coefficients $k'$ and $k''$ in each band. 
    For example, given an airmass $X$, the observed $i$ and $z$ band magnitudes of
    a standard star are related to the tabulated, intrinsic magnitudes $z_0$ and
    $i_0$ by a set of linear equations
    \begin{eqnarray}
    	\text{z} &= \text{z}_0 + k_{\text{z}}' \cdot X + k_{\text{z}}'' (\text{b} - \text{v}) \cdot X \label{eq:z} \\
    	\text{i} &= \text{i}_0 + k_{\text{i}}' \cdot X + k_{\text{i}}'' (\text{b} - \text{v}) \cdot X    \label{eq:i}
    \end{eqnarray}
    The first order extinction term $k'$ accounts for the decrease in a star's 
    observed flux with airmass. The inclusion of a second order coefficient $k''$ 
    accounts for the fact that the observed flux of blue stars decreases faster 
    than red stars as they approach the horizon.

    To measure the second-order extinction, observations are taken of a red and 
    blue star over a wide airmass range. The second-order extinction in each 
    band can then be found by fitting for the difference in magnitude between 
    the two stars.
    \begin{eqnarray}
	    \Delta \text{z} &= k''_{\text{z}} \Delta(\text{b}-\text{v}) \cdot X + \Delta \text{z}_0 \\
	    \Delta \text{i} &= k''_{\text{i}} \Delta(\text{b}-\text{v}) \cdot X + \Delta \text{i}_0
    \end{eqnarray}
    Using the resulting value for $k''$, the first order extinction coefficient 
    is then found by fitting Equations \ref{eq:z} and \ref{eq:i}. Although this 
    method does account for a first order airmass dependence, it does not 
    directly account for any nonlinear effects. This is to say it does not 
    account for parts of the atmospheric transmission having a nonlinear airmass 
    dependence.

    \begin{figure*}[t]
        \includegraphics[width=\textwidth]{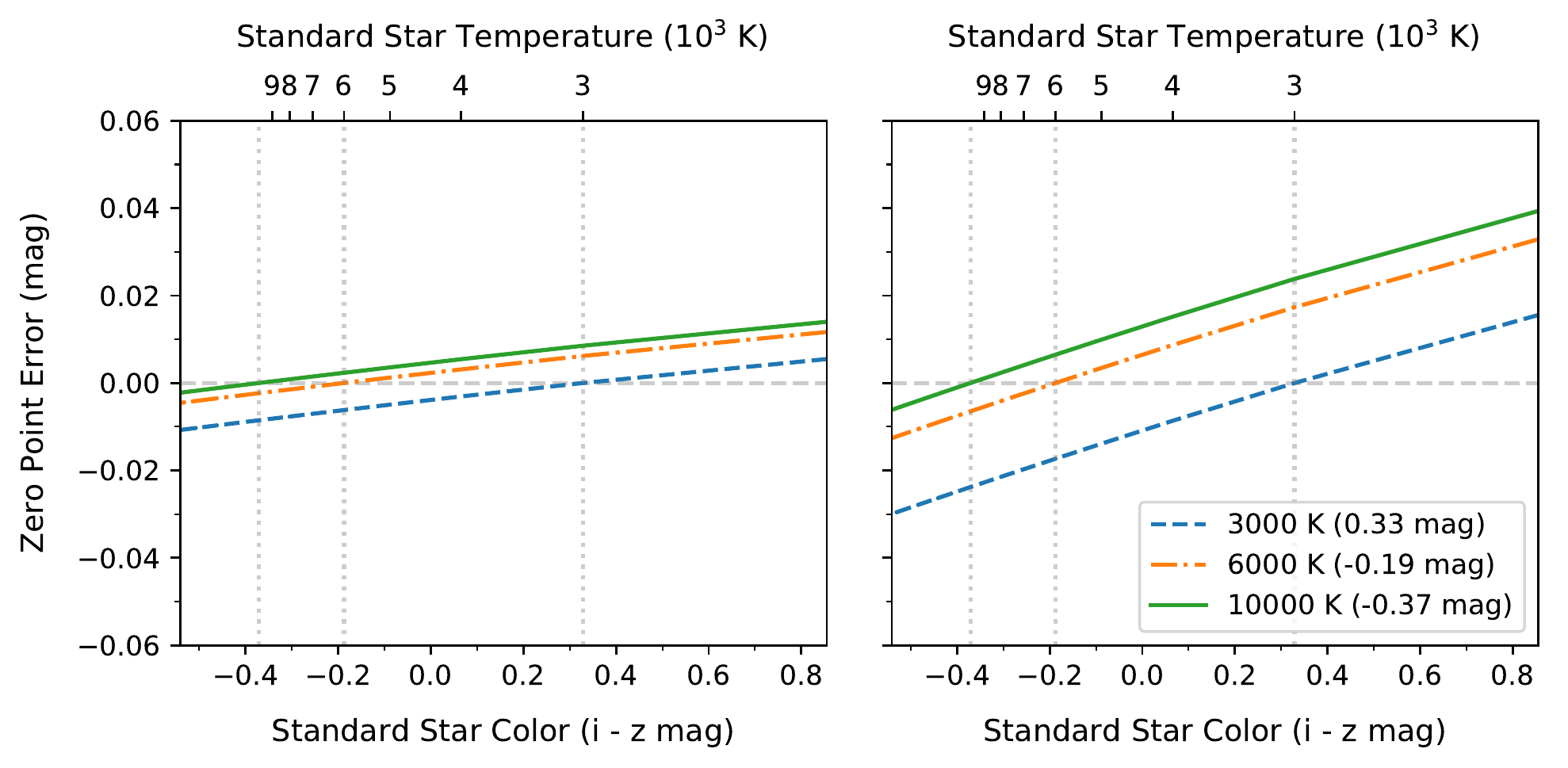}
        \caption{Correcting photometric observations using tabulated values of a 
                 standard star introduces residual error in the magnitudes of 
                 other stars with different spectral types. The residual error 
                 in $z$ band photometric zero point due to absorption by 
                 precipitable water vapor is shown for three black bodies at 
                 $3,000$ (M type), $6,000$ (G type), and $10,000$ K (A type). 
                 Results are shown as a function of the color 
                 of the reference star used to calculate the zero point. Error 
                 values are shown for a PWV column density of 5 (left) and 30 mm 
                 (right).}
        \label{fig:zp_error}
    \end{figure*}

    For a PWV column density at zenith PWV$_{z}$, the column density along the
    line of sight is given by
    \begin{equation} \label{eq:pwv_los}
        \text{PWV}_{\rm los} = \text{PWV}_{z} \cdot X
    \end{equation}
    However, due to saturation, not all absorption features scale linearly with
    PWV concentration -- some features saturate at relatively low concentrations
    ($<10$~mm). Thus a linear function of airmass and color is not sufficient to 
    describe the atmospheric transmission from PWV.

    Figure~\ref{fig:black_body} details the error introduced by considering PWV 
    absorption averaged over a bandpass versus the actual absorption spectrum. 
    Because atmospheric absorption varies with wavelength, it affects stars 
    differently depending on their spectral type. This means that variations in 
    the spectral types of photometric standards used to correct an image 
    introduce errors in the magnitudes of observed targets. This effect is more 
    pronounced for higher airmass due to the increased PWV along the line of sight, 
    and is an important consideration for KPNO where $\text{PWV}_{z}$
    exceeds $20$ mm over 13\% of the time.

    Demonstrated in Figure \ref{fig:zp_error}, when using a type A star to 
    correct cooler G or M type stars, spectral variations between stars used in 
    the atmospheric correction can introduce errors as large as $-0.02$ mag. 
    This error is particularly important when performing high accuracy photometry
    to 1\% or better. An alternative is to correct photometric observations using 
    atmospheric models.

    For an atmospheric transmission $T(\lambda)$, the photometric 
    correction for an object with a spectral energy distribution $S(\lambda)$ is 
    given by
    \begin{equation} \label{eq:atm_correction}
    C = \frac{\int_{\lambda_i}^{\lambda_j} S(\lambda) \cdot T(\lambda) \, d\lambda}
             {\int_{\lambda_i}^{\lambda_j} S(\lambda) \, d\lambda}
    \end{equation}
    where the integration bounds are defined by the wavelength range of the 
    photometric bandpass. Using atmospheric models, measurements of the PWV 
    column density are used to determine $T(\lambda)$  at a given date, time, 
    and airmass. If tabulated values for $S(\lambda)$ are not available, 
    spectral templates are used instead. For example, the SED of a star is well 
    estimated by its color, due to the strong relationship between stellar 
    spectral type and intrinsic color.

    \subsection{Use of GPS at Kitt Peak} \label{ssec:gps_at_kitt_peak}
    In March of 2015, we installed SuomiNet connected weather station on top of 
    the WIYN 3.5 meter telescope building at Kitt Peak National Observatory. In 
    addition to a GPS receiver, the station includes barometric, temperature, 
    and wind speed sensors. SuomiNet compiles measurements from its affiliated 
    weather stations at thirty minute intervals. These semi-hourly measurements, 
    in addition to the local PWV column density along zenith, are then released 
    publicly on an hourly basis.

    In order to prevent equipment damage, the weather station at Kitt Peak is 
    powered down during lightning storms. This creates gaps in the available 
    SuomiNet data for Kitt Peak. Additionally, the barometric sensor was
    malfunctioning in 2016 from January through March, so we ignore any SuomiNet 
    data published for Kitt Peak during this time period. The sensor has since 
    been repaired, but occasionally records a non-physical drop in pressure. We
    disregard these measurements by ignoring any meteorological measurements 
    taken for Kitt Peak with a pressure below 775~mbar.

\begin{figure*}
        \subfigure[]{
            \label{fig:gps_locations}
            \epsscale{.6}
            \includegraphics[width=0.45\textwidth]{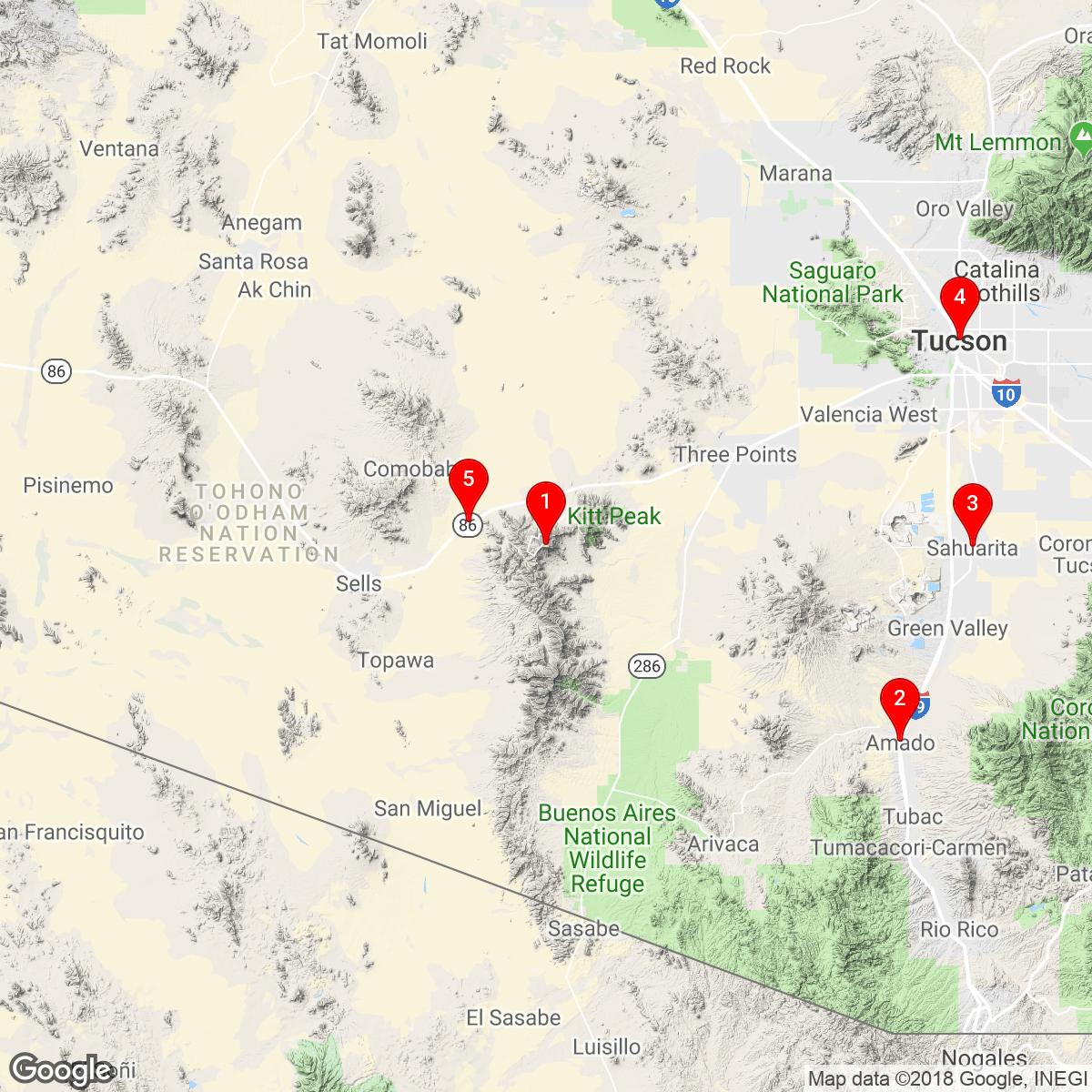} 
        } 
        \hspace{0.05\textwidth}
        \subfigure[]{
            \includegraphics[width=0.45\textwidth]{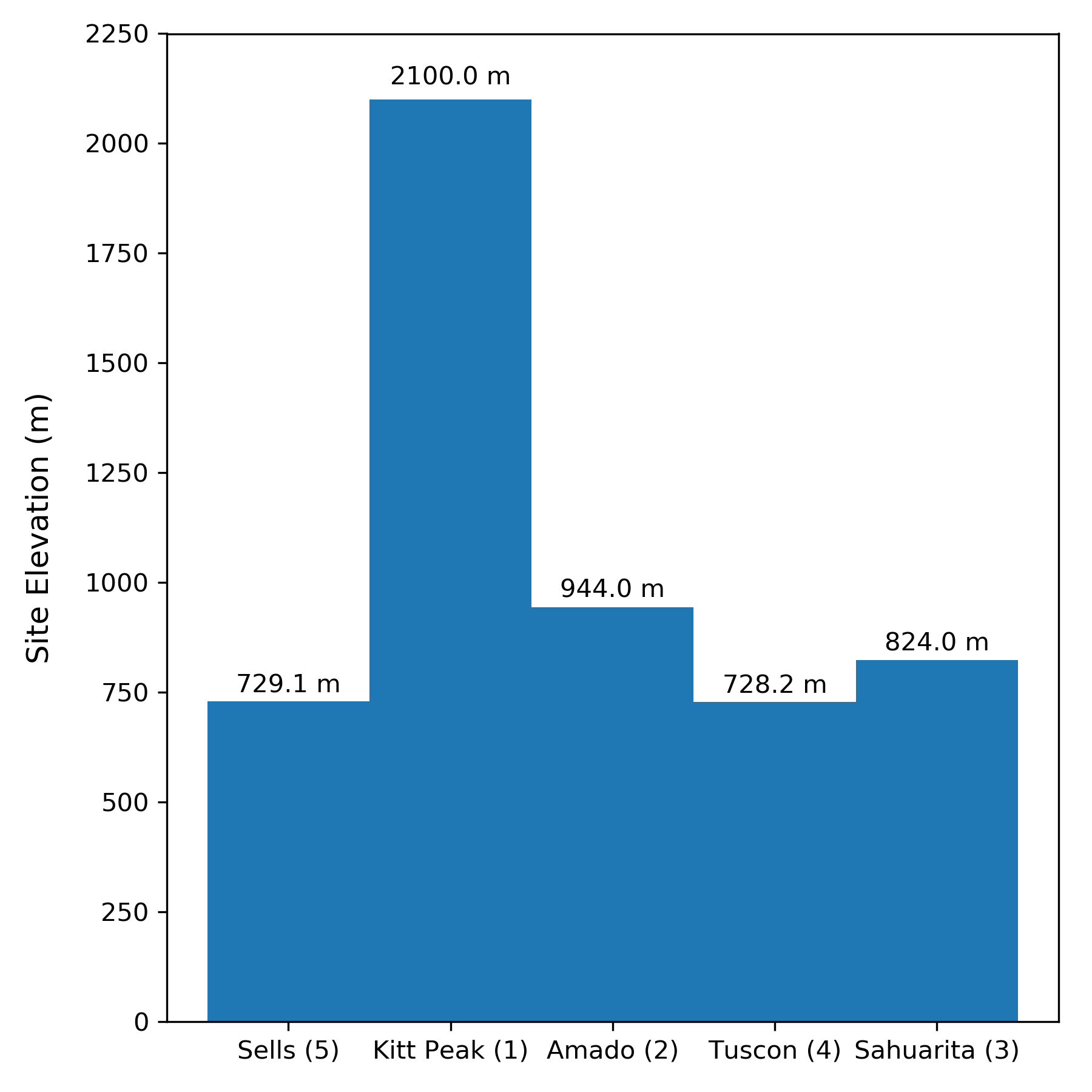}
        }
        \caption{The \pwvkpno\ package uses PWV measurements for five locations 
                 within a 45 mile radius of Kitt Peak. Shown on the left, these
                 locations include Kitt Peak (1), Amado (2), Sahuarita (3),
                 Tucson (4), and Sells (5) Arizona. The elevation of each location
                 is shown on the right.}
    \end{figure*}

\begin{figure*}
        \includegraphics[width=\textwidth]{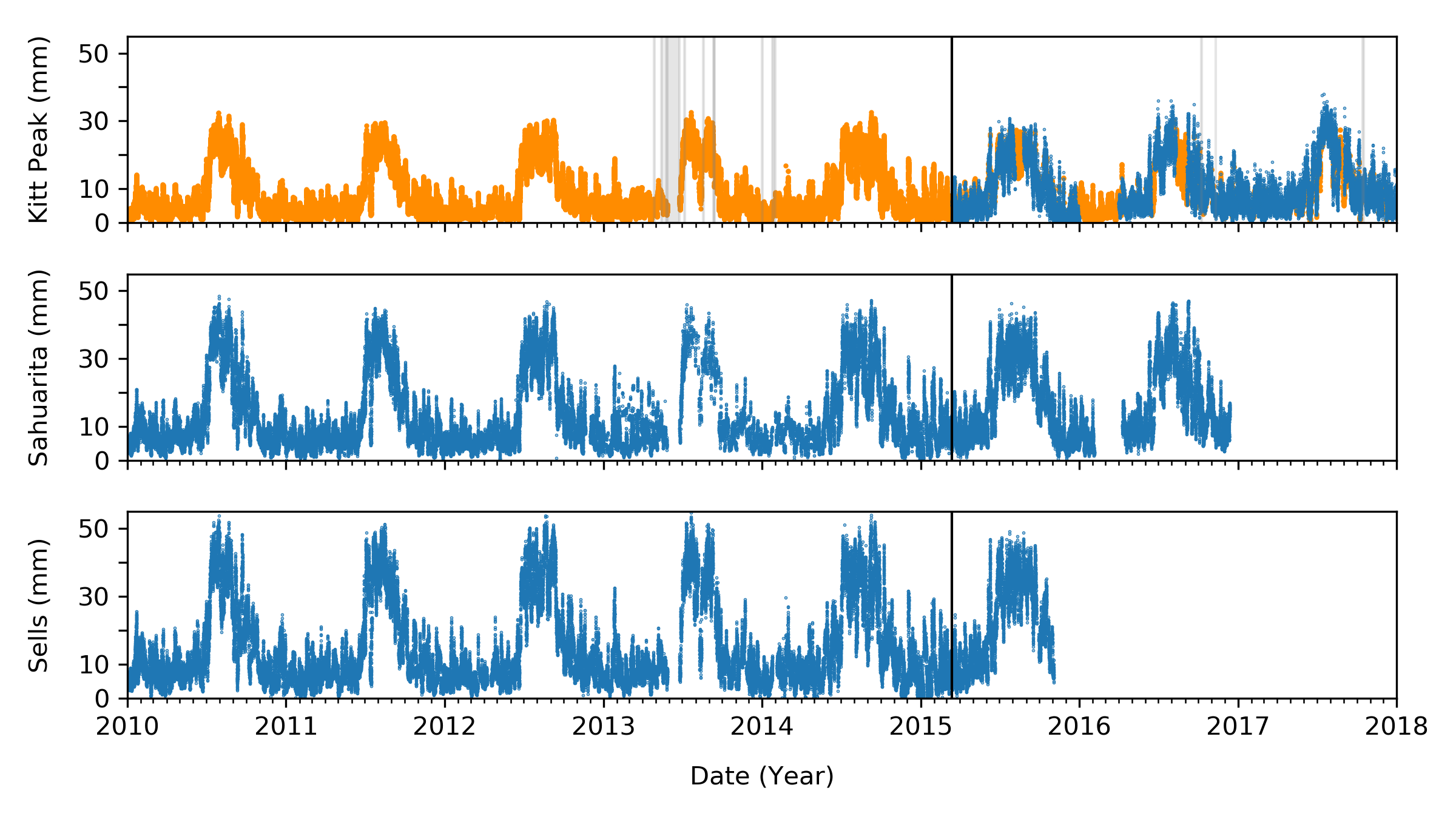}
        \caption{Measurements of precipitable water vapor (PWV) from the 
                 SuomiNet project from 2010 onward. From top to bottom, SuomiNet 
                 measurements for Kitt Peak National Observatory, Sahuarita AZ, 
                 and Sells AZ (Blue). The modeled PWV level at Kitt peak is 
                 shown in Orange. Periods of one day or longer where there are 
                 no modeled PWV values are shown in the top panel in grey. The geographic 
                 proximity of these locations means that the primary difference 
                 in PWV between locations is due to differences in altitude. 
                 Measurements taken at Kitt Peak National Observatory begin in 
                 March of 2015.}
        \label{fig:suomi_data}
    \end{figure*}

    In order to determine the PWV level during periods without SuomiNet data, 
    measurements from other nearby receivers can be used to model the PWV level 
    at Kitt Peak. This model can also be used for times before the Kitt Peak
    receiver was installed. In addition to data taken at Kitt Peak, the \pwvkpno\ 
    package uses measurements from four other receivers within a 45 mile radius
    at varying levels of altitude. 
    This includes receivers located at Amado (AMAZ), Sahuarita (P014), Tucson
    (SA46), and Sells (SA48) Arizona. The location of these receivers is shown 
    in Figure \ref{fig:gps_locations}, with SuomiNet measurements for Kitt Peak, 
    Amado, and Sells shown in Figure \ref{fig:suomi_data}.
    
    \begin{figure*}
        \epsscale{.95}
        \setlength{\belowcaptionskip}{12pt}
        \hspace*{-1cm} 
        \plotone{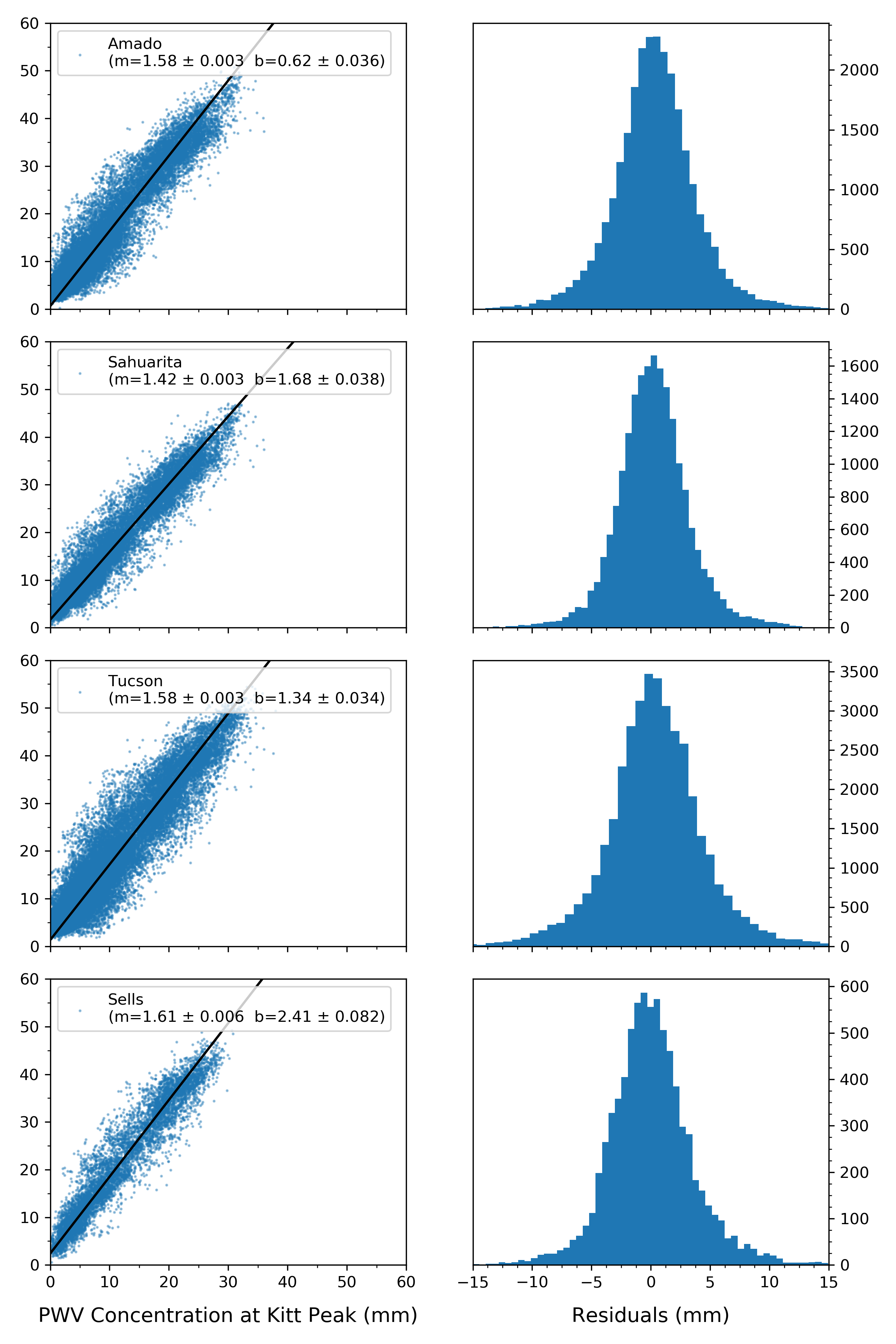}
        \caption{Linear fits to measurements of precipitable water vapor (PWV) 
                 column density taken at four different locations versus
                 simultaneous measurements taken at Kitt Peak. Each row
                 corresponds to a different location being compared against
                 Kitt Peak, with measurements shown on the left  and binned
                 residuals shown on the right. The slope ($m$) and y-intercept ($b$)
                 is shown for each fit. The correlation in PWV column
                 density between different sites allows the PWV column density
                 at Kitt Peak to be modeled using measurements from other
                 locations.}
        \label{fig:site_correlations}
    \end{figure*}

    Note that the PWV level at each location follows the same seasonal trend, 
    but the PWV concentration at Kitt Peak tends to be lower. Since each of 
    the chosen receivers are geographically close together, variations in PWV 
    between Kitt Peak and the four supplementary locations are predominantly 
    caused by differences in altitude. Shown in Figure 
    \ref{fig:site_correlations}, the PWV level at each location can be related 
    to the PWV level at Kitt Peak by applying a linear fit. Each fit is able to 
    predict the PWV column density at Kitt Peak to a precision of 1~mm plus 10\% 
    of the predicted value.
    
    For times when SuomiNet data is unavailable for Kitt Peak, each of the 
    linear fits are used to estimate the PWV column density at Kitt Peak. The 
    resulting estimations are then averaged and used to supplement data taken by 
    the Kitt Peak weather station. This full data set provides a model for the 
    PWV column density at zenith over time.

    To determine the PWV column density for a specific date and time, \pwvkpno\
    first determines the concentration along zenith by interpolating from the
    supplemented PWV data. The PWV column density along the line of site is then
    calculated using Equation \ref{eq:pwv_los}. Using this value, \pwvkpno\ is
    able to determine the atmospheric transmission using a set of tabulated
    MODTRAN models.

%% file: features.tex
\section{Features and Use of pwv\_kpno} \label{sec:features_and_use}
    The \pwvkpno\ package provides access to models for the atmospheric 
    transmission due to PWV at any location within the SuomiNet GPS network.
    However, the package is configured by default to return models for Kitt Peak
    National Observatory. We here demonstrate the features of \pwvkpno\ using
    the default model for Kitt Peak and further discuss modeling custom
    sites in Section \ref{sec:other_locations}
    
    \pwvkpno\ is registered with the Python Package Index and is compatible
    with both Python 2.7 and 3.5 through 3.7. Using PWV measurements published
    by the SuomiNet project, the package is able to determine the atmospheric
    transmission between $3,000$ and $12,000$ \r A. The package also provides
    methods for the automated retrieval and processing of published SuomiNet data.
    
    \subsection{Accessing PWV Data} \label{ssec:accessing_pwv}
    In order to model the atmospheric transmission for a given date and time, 
    \pwvkpno\ requires there to be SuomiNet data stored on the user's local 
    machine. Each package release contains the necessary data to return models 
    for Kitt Peak from 2010 through the end of the previous year. This data is
    automatically included when installing the package.

    Access to tabulated PWV data and modeling of the PWV transmission function 
    is provided by the \code{pwv\_atm} module. A list of years that have been 
    downloaded from SuomiNet to the user's local machine can be retrieved
    using the \code{downloaded\_years} method.
    
    \begin{lstlisting}
>>> from pwv_kpno import pwv_atm
>>> pwv_atm.downloaded_years()

  [2010, 2011, 2012, 2013, 2014,
        2015, 2016, 2017]
    \end{lstlisting}
    The returned list includes all years for which any amount of data has been 
    downloaded.

    In order to update the locally stored data, \pwvkpno\ can be used to 
    automatically retrieve and processes new data from SuomiNet. This is 
    achieved using the \code{update\_models} method.
    \begin{lstlisting}
>>> pwv_atm.update_models()

    [2017, 2018]
    \end{lstlisting}
    Here the returned list includes any years for which new data was 
    downloaded. By default, the function will download all published data for 
    any years not currently present on the local machine. In addition, it will 
    also download data for the most recent year that is locally available. This 
    method ensures there are no years with incomplete measurements in the 
    locally available data. If desired, the user can alternatively specify a 
    specific year to download from 2010 onward.

    In addition to downloading data for Kitt Peak, the \code{update\_models} 
    method also downloads measurements taken at the four supplementary 
    locations shown in Figure \ref{fig:gps_locations}. Each time the method is 
    run, a new set of linear fits is created to describe the PWV concentration 
    at Kitt Peak as a function of the PWV concentration at each supplementary 
    location. These new fits are then used to recreate the entire supplemented PWV 
    model for Kitt Peak. The error in PWV modeled using each of these fits 
    is taken as the standard deviation of that fit's residuals.

    Users can access the locally available SuomiNet data using the 
    \code{measured\_pwv} method. Results are returned as an Astropy table 
    \citep{astropy} and can be independently filtered by year, month, day, and 
    hour. 
    \begin{lstlisting}
>>> pwv_atm.measured_pwv(
        year=2016, month=11, day=14)

         date      KITT KITT_err P014 ...
         UTC        mm     mm     mm  ...
  ---------------- ---- -------- ---- ...
  2016-11-14 00:15  4.7    1.025  6.9 ...
  2016-11-14 00:45  4.3    1.025  6.7 ...
  2016-11-14 01:15  3.9    0.925  6.7 ...
               ...  ...      ...  ... ...
    \end{lstlisting}
    Excluding the date column, each column is labeled using the SuomiNet 
    identification codes for the GPS receivers. 
        
    \pwvkpno\ also provides access to the modeled PWV column density at Kitt
    peak via the \code{modeled\_pwv} method. As in the previous example,
    these results can also be filtered independently by year, month,
    day, and hour.
    \begin{lstlisting}
>>> pwv_atm.modeled_pwv(
        year=2016, month=11, day=14)

         date      pwv pwv_err
         UTC        mm    mm  
  ---------------- --- -------
  2016-11-14 00:15 4.7   1.025
  2016-11-14 00:45 4.3   1.025
  2016-11-14 01:15 3.9   0.925
               ... ...     ...
    \end{lstlisting}

    \subsection{Modeling the Atmosphere} \label{ssec:modeling_atm}
    For a known PWV column density, the package provides access to the modeled 
    atmospheric transmission via the \code{trans\_for\_pwv} function. This 
    method returns the modeled transmission function as an Astropy table 
    with wavelengths ranging from $3,000$ to $12,000$ \r A. For example, given 
    a PWV column density of $13.5$ mm: 
    \begin{lstlisting}
>>> pwv_atm.trans_for_pwv(13.5)

  wavelength transmission 
   Angstrom                
  ---------- ------------ 
     3000.00 0.9999999916 
     3000.05 0.9999999916 
     3000.10 0.9999999916 
         ...          ... 
    \end{lstlisting} 
    Atmospheric models can also be accessed for a given
    datetime and airmass using the function 
    \code{trans\_for\_date}. 
    \begin{lstlisting}
>>> from datetime import datetime
>>> import pytz
>>>
>>> obsv_date = datetime(
>>>     year=2013, month=12, day=15,
>>>     hour=5, minute=35, tzinfo=pytz.utc)
>>>
>>> pwv_atm.trans_for_date(
        date=obsv_date, airmass=1.2)

  wavelength transmission transmission_err
   Angstrom                                  
  ---------- ------------ ----------------
     3000.00 0.9999999916 1.7305648359e-08
     3000.05 0.9999999916 1.7305648359e-08
     3000.10 0.9999999916 1.7305648359e-08
         ...          ...              ...
    \end{lstlisting}
    
    If \pwvkpno\ does not have any supplemented SuomiNet data within a day
    of the requested datetime, an exception is raised. Both the
    \code{trans\_for\_pwv} and \code{trans\_for\_date} functions 
    determine the atmospheric transmission by returning a set of MODTRAN 
    transmission models. 

    \subsection{Modeling a Black Body} \label{ssec:modeling_blackbody}

    The \code{blackbody\_with\_atm} module provides functions for modeling the 
    effects of PWV absorption on a black body SED. For example, consider a 
    black body at \mbox{$8,000$ K} under the effects of atmospheric absorption 
    due to $15$ mm of PWV. For a given array of wavelengths in angstroms, the 
    \code{sed} method returns the corresponding spectral energy distribution. 
    \begin{lstlisting}
>>> from pwv_kpno import blackbody_with_atm \
>>>     as bb_atm
>>>
>>> temp = 8000
>>> wavelength = np.arange(7000, 10000, 100)
>>> pwv = 15
>>>
>>> sed = bb_atm.sed(temp, wavelength, pwv)
    \end{lstlisting}
    The SED from the above example can be seen in Figure \ref{fig:black_body}. 
    If desired, the SED of a black body without atmospheric effects can also be 
    achieved by specifying a PWV column density of zero.
    
    Using the \code{magnitude} function, users can determine the magnitude 
    of a black body in a given band. For example, in the $i$ band, which ranges 
    from $7,000$ to $8,500$ \r A, the AB magnitude of a black body is found by 
    running 
    \begin{lstlisting}
>>> band = (7000, 8500)
>>> mag = bb_atm.magnitude(temp, band, pwv)
    \end{lstlisting}
    Here the $i$ band is treated as a top-hat function, however, the
    \code{magnitude} function also accepts \code{band} as a two dimensional 
    array specifying the wavelength and response function of a real-world
    band. As in the previous example, the magnitude of a black body without 
    the effects of atmospheric absorption can be found by specifying a PWV 
    level of zero.

%% file: other_locations.tex
\section{Modeling Other Locations} \label{sec:other_locations}
    By default, \pwvkpno\ provides models for the PWV transmission 
    function at Kitt Peak National Observatory. However, \pwvkpno\ also
    provides atmospheric modeling for user customized locations. 
    Modeling multiple locations is handled by the \code{package\_settings} 
    module, and allows modeling at any location with a SuomiNet connected 
    GPS receiver.
    
    Each site modeled by \pwvkpno\ is represented by a unique configuration
    file. Using the \code{ConfigBuilder} class, users can create customized
    configuration files for any SuomiNet site. As an example, we create a
    new model for the Cerro Tololo Inter-American Observatory (CTIO) near
    La Serena, Chile.
    \begin{lstlisting}
>>> from pwv_kpno.package_settings import \
>>>     ConfigBuilder
>>>
>>> new_config = ConfigBuilder(
        site_name='cerro_tololo',
        primary_rec='CTIO',
        sup_rec=[],
        wavelength=custom_wavelengths,
        cross_section=custom_cross_sections
)

>>> new_config.save_to_ecsv(
        './cerro_tololo.ecsv')
    \end{lstlisting} 

    Here \code{site\_name} specifies a unique identifier for the site being modeled, \code{primary\_rec} is the
    SuomiNet ID code for the GPS receiver located at the modeled site, and \code{sup\_rec} is a list of SuomiNet ID 
    codes for nearby receivers used to supplement measurements taken by the primary receiver. Unlike the default model 
    for KPNO, there are no additional receivers near the CTIO and so \code{sup\_rec} in this example is left empty (the 
    default value). By default, \pwvkpno\ models use MODTRAN estimates for the wavelength dependent cross section of 
    H$_2$O from 3,000 to 12,000 \AA. The optional \code{wavelength} and \code{cross\_section} arguments allow a user 
    to customize these cross sections in units of Angstroms and cm${^2}$ respectively.
    
    If desired, users can specify custom data cuts on SuomiNet data used by the package. Data cuts are defined using a 
    2d dictionary of boundary values. The first key specifies which receiver the data cuts apply to. The second key 
    specifies what values to cut. Following SuomiNet's naming convention, values that can be cut include PWV \
    (\code{"PWV"}), the PWV error (\code{"PWVerr"}), surface pressure (\code{"SrfcPress"}), surface temperature 
    (\code{"SrfcTemp"}), and relative humidity (\code{"SrfcRH"}). For example, if we wanted to ignore measurements
    taken between two dates, we can specify those dates as UTC timestamps and run
    \newline
    \newline
    \newline
    \begin{lstlisting}
>>> data_cuts = {'CTIO': 
        {'SrfcPress': [
            [time_start, time_end]
        ]}
    }   

>>> new_config = ConfigBuilder(
        site_name='cerro_tololo',
        primary_rec='CTIO',
        data_cuts=data_cuts)
    \end{lstlisting}

    Once a configuration file has been created, it can be permanently added to the 
    locally installed \pwvkpno\ package by running
    \begin{lstlisting}
>>> from pwv_kpno.package_settings import \
>>>     settings
>>>
>>> settings.import_site_config(
    './cerro_tololo.ecsv')
    \end{lstlisting}
    This command only needs to be run once, after which \pwvkpno\ will retain the new model on disk, even in between
    package updates. The package can then be configured to use the new model by running
    \begin{lstlisting}
>>> settings.set_site('cerro_tololo')
    \end{lstlisting}  
 
    After setting \pwvkpno\ to a model a specific site, the package will return atmospheric models and PWV data 
    exclusively for that site. It is important to note that this setting is not persistent. When \pwvkpno\ is first 
    imported into a new environment the package will always default to using the standard model for Kitt Peak, and the 
    above command will have to be rerun.
    
    A complete summary of package settings can be accessed using attributes of the \code{settings} object.
    \begin{lstlisting}
>>> settings.set_site('kitt_peak')
>>> print(settings.site_name)
  kitt_peak
    
>>> print(settings.available_sites)
  ['kitt_peak', 'cerro_tololo']
    
>>> print(settings.receivers)
  ['AZAM', 'KITT', 'P014', 'SA46', 'SA48']
    
>>> print(settings.primary_rec)
  KITT

>>> print(settings.supplement_rec)
  ['AZAM', 'P014', 'SA46', 'SA48']
    \end{lstlisting} 
    The configuration file for the currently modeled location can be exported by running
    \begin{lstlisting}
>>> settings.export_site_config(
    './current_site_name.ecsv')
    \end{lstlisting}

%% file: validation.tex
\section{Validation} \label{sec:validation}
    \begin{figure*}
        \epsscale{.9}
        \plotone{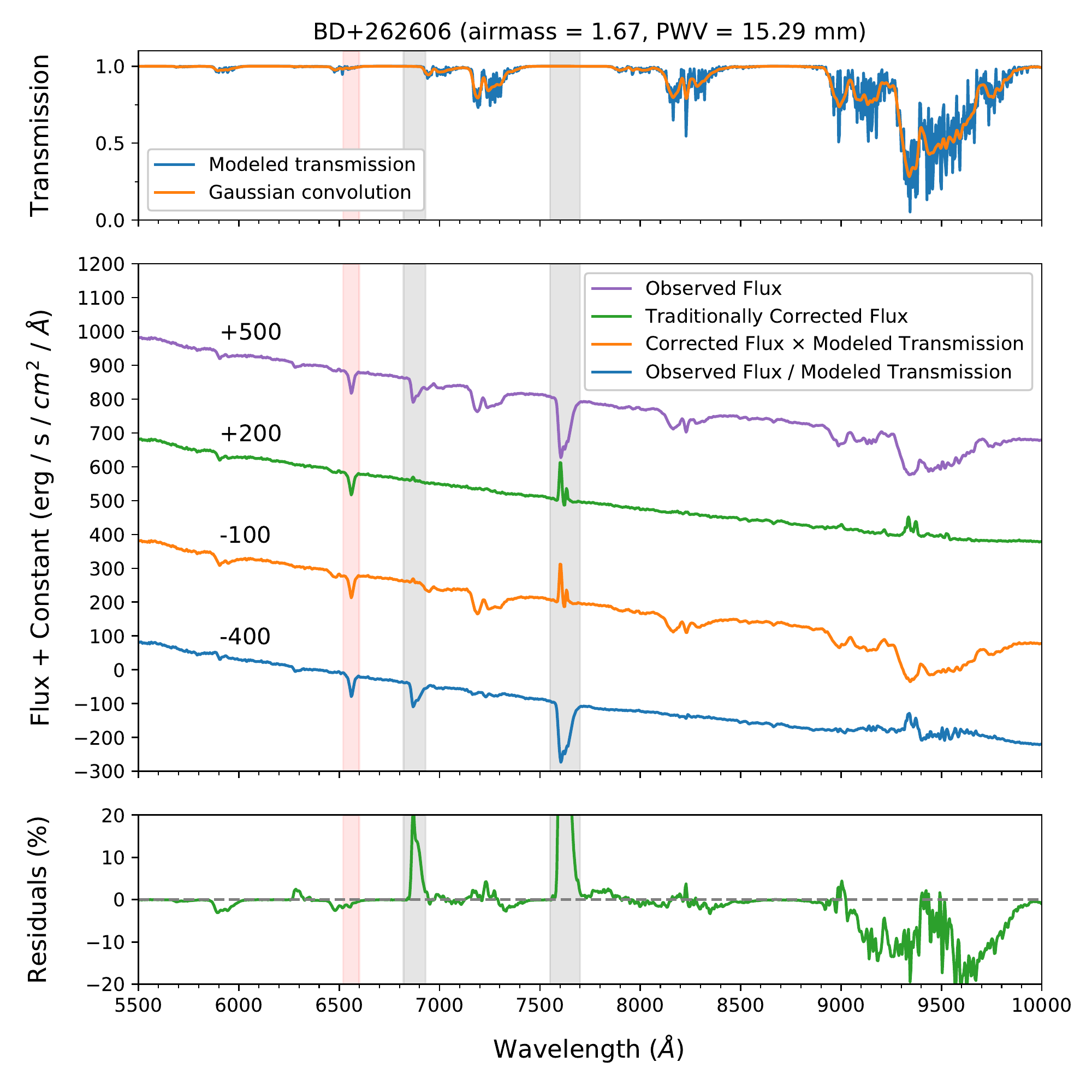}
        \caption{Observations of BD+262606 were taken using the R. C.
        Spectrograph at Kitt Peak National Observatory. The top panel
        demonstrates the modeled PWV transmission function at the time of
        observation (blue) smoothed by a Gaussian kernel (orange). The middle
        panel shows the observed spectrum (purple) and the spectrum
        corrected using catalog values (green). These are compared against the
        catalog corrected spectrum multiplied by the modeled transmission (orange)
        and the observed spectrum divided by the smoothed transmission (blue).
        Residuals between the catalog corrected and model corrected spectrum
        are shown in the bottom panel. H$_\alpha$ lines are highlighted in
        red and O$_2$ lines in grey.}
        \label{fig:spectral_correction}
    \end{figure*}

    From 2010 September 16th through September 20th, an observation
    run was performed on 18 standard stars using the R.C. spectrograph
    on the Mayall 4m telescope. To reduce flux
    loss due to atmospheric dispersion, the spectrograph was configured to
    use a wide 7\arcsec\ slit. Observations were recorded between 5,500 and
    10,200 \r A with an average dispersion of 3.4 \r A per pixel. Seeing for
    all observations varied between 1 and 2\arcsec. 
    
    As an example, Figure 7 shows the SED of BD+262606 observed at an
    airmass of 1.67. To flux-calibrate the observed spectrum, low-airmass
    observations were taken of BD+17 4708 each night. This minimized the
    introduction of additional telluric effects in the calibrated
    spectrum. To correct the observed spectrum for atmospheric effects, the
    absorption in the standard star was scaled to match the airmass of the
    other observations following the prescription of \citet{wade88}.
    
    Note that the atmospheric models used by \pwvkpno\ do not directly 
    account for the smoothing that occurs in observed spectra due to a 
    spectrograph's spectral resolution function. As a result, directly dividing 
    the observed spectra and modeled transmission will produce a very high,
    unphysical flux for wavelengths where the transmission function is saturated.
    To account for any saturated features, the modeled transmission is first
    binned to approximately match the observed spectrum's resolution. The
    transmission is then smoothed further using a Gaussian kernel.
    
    To correct for atmospheric effects using the \pwvkpno\ package, the
    observed spectrum is divided by the smoothed PWV 
    transmission function. We note that the observed spectrum was taken
    before a GPS receiver was installed at Kitt Peak. This means that
    no direct PWV measurements are available for the time of observation,
    and we instead determine the modeled PWV transmission using
    measurements from GPS receivers on the surrounding desert floor.
    
    In the model-corrected spectrum, the 
    absorption feature at 6,550 \r A is an H$\alpha$ line intrinsic to the 
    observed spectrum. Furthermore, the absorption features at 6,875 and 7,650 
    \AA\ are caused by O$_2$ absorption. Since \pwvkpno\ only provides models 
    for the PWV absorption, these two features remain uncorrected. Given that
    there are no emission lines relative to the continuum, the feature at 9350 
    \AA\ is categorized as an unidentified artifact from the reduction process.
     
    Corrections for the PWV absorption features agree reasonably well between
    the catalog and model corrected spectrum. The largest deviations between the 
    corrected spectrum occur redward of 9,000 \AA. Some of these deviations can 
    be attributed to cloudy observation conditions, creating large spatial
    and time variations in the PWV concentration along the line of sight
    \citep{querel14}. However, correcting this feature is also difficult since
    it is in fact a number of thin, saturated lines that have been blended 
    together. Overall we find that the model struggles to correct the observed
    spectrum past 9,000 \r A, but performed well enough overall to be used to 
    satisfactorily correct photometric observations.

%% file: demonstration.tex
\section{Package Demonstration} \label{sec:package_demo}    
    The \pwvkpno\ package can be used to correct both spectrographic and photometric observations. 
    As an example, we use the \pwvkpno\ package to determine the atmospheric correction presented in
    Figure \ref{fig:spectral_correction}. We also demonstrate how to calculate the photometric correction
    factor defined in Equation \ref{eq:atm_correction} for a black body.
    
    \subsection{Correcting Spectra} \label{ssec:correcting_spectra}
    Spectrographic observations are corrected by dividing observed spectra by the modeled atmospheric
    transmission function. To account for the spectral resolution function of the observing spectrograph, the
    modeled transmission is first binned to approximately match the observed spectra's 
    resolution. Depending on the resolution of the observation, further smoothing can then be performed using a 
    Gaussian kernel. Assume that the observed wavelength and flux values are stored in equal length arrays 
    \code{obs\_wavelength} and \code{obs\_flux} respectively. Using the date, time, and airmass of the observation,
    the binned transmission function is found by running
    \begin{lstlisting}
>>> import numpy as np
>>>
>>> resolution = 16  # Angstroms
>>> bins = np.arange(min(obs_wavelength), 
                     max(obs_wavelength + 1),
                     resolution)
>>>
>>> airmass = 1.2
>>> obs_data = datetime(2010, 09, 19, 6, 29
                        tz_info=pytz.utc)
>>> transm = pwv_atm.trans_for_date(obs_date,
                                    airmass,
                                    bins)
    \end{lstlisting}

    In order to divide the observed spectrum and modeled transmission, we linearly interpolate the binned transmission 
    to the observed wavelength values. We then apply a Gaussian smoothing using an arbitrary standard deviation of 2 
    \AA.
    
    \begin{lstlisting}
>>> from scipy.ndimage.filters import \
        gaussian_filter
>>>
>>> interp_transm = np.interp(
        obs_wavelength,
        transm['wavelength'],
        transm['transmission'])
>>> smoothed_transm = gaussian_filter(
        input=interp_transm, sigma=2)
    \end{lstlisting}
    The corrected spectrum is then given as the observed flux 
    divided by the smoothed transmission function on a wavelength by wavelength basis.
	\begin{lstlisting}
>>> corrected_spec = np.divide(
        obs_flux, 
        smoothed_transm)
	\end{lstlisting}
	
	\subsection{Correcting Photometry} \label{ssec:correcting_photometry}
    The \pwvkpno\ package can also be used to correct photometric observations of objects with a known spectral
    type. To do so, it 
    is necessary to evaluate Equation \ref{eq:atm_correction}. Note that the product in the numerator $S(\lambda) \cdot 
    T(\lambda)$ represents the SED under the influence of atmospheric effects, while $S(\lambda)$ in the denominator 
    represents the intrinsic SED. For a black body observed in the $i$ band, these values can be found as
	\begin{lstlisting}
>>> # S(lambda) * T(lambda)
>>> sed_with_atm = bb_atm.sed(
>>>     sed_temp, i_band, pwv)  
>>>
>>> # S(lambda)
>>> intrinsic_sed = bb_atm.sed(
>>>     sed_temp, i_band, 0)
	\end{lstlisting}
	In practice the SED of a photometrically observed object may not be available. In such a case it is sufficient 
	to use spectral templates instead. For example, the SED of a star can be reasonably well parametrized by its 
	observed color.
	
	Using the above results, we evaluate Equation \ref{eq:atm_correction} by performing trapezoidal integration with 
	the Numpy package. 
	\begin{lstlisting}
>>> numerator = np.trapz(sed_with_atm, i_band)
>>> denominator = np.trapz(
>>>    intrinsic_sed, i_band)
>>>
>>> photo_corr = np.divide(
>>>     numerator, denominator)
	\end{lstlisting}
	The corrected photometric flux of the black body is then found by dividing the observed flux by the correction 
	factor \code{photo\_corr}.
    

%% file: conclusion.tex
\section{Conclusion and Future Work} \label{sec:conclusion}
    Atmospheric transmission in the near-infrared is highly dependent on the 
    column density of precipitable water vapor. By measuring the delay in GPS 
    signals through the atmosphere, initiatives such as the SuomiNet project 
    provide accurate water vapor measurements for multiple, international 
    locations. Through the use of atmospheric models, these measurements provide 
    a means for determining the atmospheric transmission due to precipitable 
    water vapor at each location.
    
    Current methods for removing atmospheric effects commonly rely on fitting 
    for a set of extinction coefficients. Unfortunately, this method does not 
    capture the complex nature of the atmospheric transmission function.  
    When calibrating a photometric image, this introduces errors due to spectral 
    variations of the stars used to determine the extinction coefficients. 
    Atmospheric modeling has the potential to provide an alternative 
    that is not influenced by spectral differences.

    The Python package \pwvkpno\ provides models for the atmospheric transmission 
    due to H$_2$O at user specified sites. For a given date, time, and airmass,
    the package uses 
    measurements from the SuomiNet project to determine the corresponding PWV 
    column density along the line of sight. By using a set of MODTRAN models,
    the resulting concentration is then used to determine the PWV
    transmission function between $3,000$ and $12,000$ \r A.
    
    Future work is planned by the primary author to further explore the 
    relationship between PWV measured by geographically separated GPS receivers. 
    Measurements from two, geographically close receivers can be related 
    by a linearly fitting the PWV concentration measured at both sites. However, 
    this linear relationship does not capture the intrinsic scatter of the 
    measured data. Additional models will be explored that take into account 
    simultaneous temperature, pressure, and relative humidity measurements to 
    improve the ability to model the PWV relationship between GPS receivers.

%% file: acknowledgments.tex
\acknowledgments
\textbf{Acknowledgments:}
    This work was supported in part by the US Department of Energy Office of Science under DE-SC0007914.

    This work is based in part on observations taken at Kitt Peak National Observatory, National Optical Astronomy 
    Observatory (NOAO Prop. IDs: 2011B-0482 and 2012B-0500; PI: Wood-Vasey), which is operated by the Association of 
    Universities for Research in Astronomy (AURA) under a cooperative agreement with the National Science Foundation.
    
    We acknowledge Jessica Kroboth for investigating the relationship between PWV concentrations measured at Kitt Peak and 
    on the desert floor. 
    
    We acknowledge Abhijit Saha and David Burke for assisting in the reduction of spectral observations taken at Kitt
    Peak National Observatory. 

\software{Python\footnote{\url{http://python.org}}, 
          AstroPy~\citep{astropy}\footnote{\url{http://www.astropy.org}},
          NumPy\footnote{\url{http://www.numpy.org}}, 
          SciPy\footnote{\url{http://www.scipy.org}}, 
          Requests\footnote{\url{http://python-requests.org}},
          Pytz\footnote{\url{https://pypi.org/project/pytz/}}
}
\clearpage

%% file: pwv_paper.bbl
\begin{thebibliography}{}
\expandafter\ifx\csname natexlab\endcsname\relax\def\natexlab#1{#1}\fi

\bibitem[{{Astropy Collaboration} {et~al.}(2013){Astropy Collaboration},
  {Robitaille}, {Tollerud}, {Greenfield}, {Droettboom}, {Bray}, {Aldcroft},
  {Davis}, {Ginsburg}, {Price-Whelan}, {Kerzendorf}, {Conley}, {Crighton},
  {Barbary}, {Muna}, {Ferguson}, {Grollier}, {Parikh}, {Nair}, {Unther},
  {Deil}, {Woillez}, {Conseil}, {Kramer}, {Turner}, {Singer}, {Fox}, {Weaver},
  {Zabalza}, {Edwards}, {Azalee Bostroem}, {Burke}, {Casey}, {Crawford},
  {Dencheva}, {Ely}, {Jenness}, {Labrie}, {Lim}, {Pierfederici}, {Pontzen},
  {Ptak}, {Refsdal}, {Servillat}, \& {Streicher}}]{astropy}
{Astropy Collaboration}, {Robitaille}, T.~P., {Tollerud}, E.~J., {et~al.} 2013,
  \aap, 558, A33

\bibitem[{{Berk} {et~al.}(2014){Berk}, {Conforti}, {Kennett}, {Perkins},
  {Hawes}, \& {van den Bosch}}]{modtran}
{Berk}, A., {Conforti}, P., {Kennett}, R., {et~al.} 2014, \procspie, 9088, 9088

\bibitem[{{Blake} \& {Shaw}(2011)}]{blake11}
{Blake}, C.~H., \& {Shaw}, M.~M. 2011, \pasp, 123, 1302

\bibitem[{Braun \& Hove(2001)}]{braun01}
Braun, J., \& Hove, T. 2001, in Proceedings of the 18th International Technical
  Meeting of the Satellite Division of The Institute of Navigation

\bibitem[{Burke {et~al.}(2010)Burke, Axelrod, Blondin, Claver, Željko Ivezić,
  Jones, Saha, Smith, Smith, \& Stubbs}]{burke10}
Burke, D.~L., Axelrod, T., Blondin, S., {et~al.} 2010, The Astrophysical
  Journal, 720, 811

\bibitem[{Burke {et~al.}(2014)Burke, Saha, Claver, Axelrod, Claver, DePoy,
  Željko Ivezić, Jones, Smith, \& Stubbs}]{burke14}
Burke, D.~L., Saha, A., Claver, J., {et~al.} 2014, The Astronomical Journal,
  147, 19

\bibitem[{{Dumont} \& {Zabransky}(2001)}]{dumont01}
{Dumont}, D.~M., \& {Zabransky}, J. 2001, in Proceedings of the Eleventh
  Symposium on Meteorological Observations and Instrumentation, 245--247

\bibitem[{{Li} {et~al.}(2016){Li}, {DePoy}, {Kessler}, {Burke}, {Marshall},
  {Wise}, {Rheault}, {Carona}, {Boada}, {Prochaska}, \& {Allen}}]{li16}
{Li}, T., {DePoy}, D.~L., {Kessler}, R., {et~al.} 2016, in Astronomical Society
  of the Pacific Conference Series, Vol. 503, The Science of Calibration, ed.
  S.~{Deustua}, S.~{Allam}, D.~{Tucker}, \& J.~A. {Smith}, 25

\bibitem[{{Nahmias} \& {Zabransky}(2004)}]{nahmias04}
{Nahmias}, M.~H., \& {Zabransky}, J. 2004, BAMS

\bibitem[{{Querel} \& {Kerber}(2014)}]{querel14}
{Querel}, R.~R., \& {Kerber}, F. 2014, in \procspie, Vol. 9147, Ground-based
  and Airborne Instrumentation for Astronomy V, 914792

\bibitem[{{The Nearby Supernova Factory} {et~al.}(2013){The Nearby Supernova
  Factory}, {Buton, C.}, {Copin, Y.}, {Aldering, G.}, {Antilogus, P.}, {Aragon,
  C.}, {Bailey, S.}, {Baltay, C.}, {Bongard, S.}, {Canto, A.}, {Cellier-Holzem,
  F.}, {Childress, M.}, {Chotard, N.}, {Fakhouri, H. K.}, {Gangler, E.}, {Guy,
  J.}, {Hsiao, E. Y.}, {Kerschhaggl, M.}, {Kowalski, M.}, {Loken, S.}, {Nugent,
  P.}, {Paech, K.}, {Pain, R.}, {P\'econtal, E.}, {Pereira, R.}, {Perlmutter,
  S.}, {Rabinowitz, D.}, {Rigault, M.}, {Runge, K.}, {Scalzo, R.}, {Smadja,
  G.}, {Tao, C.}, {Thomas, R. C.}, {Weaver, B. A.}, \& {Wu, C.}}]{TNSF13}
{The Nearby Supernova Factory}, {Buton, C.}, {Copin, Y.}, {et~al.} 2013, A\&A,
  549, A8

\bibitem[{{Vacca} {et~al.}(2003){Vacca}, {Cushing}, \& {Rayner}}]{Vacca03}
{Vacca}, W.~D., {Cushing}, M.~C., \& {Rayner}, J.~T. 2003, \pasp, 115, 389

\bibitem[{{Wade} \& {Horne}(1988)}]{wade88}
{Wade}, R.~A., \& {Horne}, K. 1988, \apj, 324, 411

\bibitem[{{Ware} {et~al.}(2000){Ware}, {Fulker}, {Stein}, {Anderson}, {Avery},
  {Clark}, {Droegemeier}, {Kuettner}, {Minster}, \& {Sorooshian}}]{ware00}
{Ware}, R.~H., {Fulker}, D.~W., {Stein}, S.~A., {et~al.} 2000, Bulletin of the
  American Meteorological Society, 81, 677

\end{thebibliography}
